\documentstyle[11pt,rafringe,twoside,epsfig]{article}

\def\edcomment#1{\iffalse\marginpar{\raggedright\sl#1\/}\else\relax\fi}
\reversemarginpar

\begin{document}
\title{VLBI Imaging of Seyfert Galaxies}
 \author{James S.~Ulvestad}
\affil{National Radio Astronomy Observatory, P.O. Box O,
Socorro, NM 87801, USA}

\begin{abstract}
The advent of the Very Long Baseline Array and its phase-referencing
capability have enabled milliarcsecond-scale imaging of radio
sources at sub-millijansky sensitivity, opening the door
to high-resolution imaging of Seyfert galaxies.  Over the last few 
years, this has led to a number of interesting new results 
that shed light on the characteristics of the inner cores
of Seyfert galaxies.  These include the following: 
(1) classical Seyfert galaxies with steep radio spectra have
a strong tendency to exhibit radio jets at milliarcsecond
scales, sometimes with considerable curvature in the jets;
(2) apparent speeds of
jets imaged at multiple epochs generally are considerably less
than $c$, often in the vicinity of $0.1c$ or less, although
there sometimes are faster motions seen after strong outbursts; and
(3) lower luminosity Seyfert galaxies have a much
stronger tendency to show flat or somewhat inverted radio spectra,
usually are unresolved on milliarcsecond scales, and probably are
dominated by a combination of low-radiative-efficiency accretion 
flows and very compact radio jets.  This paper discusses some
examples that illustrate these properties, and summarizes some of 
what we have learned about weak active galaxies on milliarcsecond
scales.
\end{abstract}

\section{Introduction}

Seyfert galaxies were first identified as a class by Seyfert (1943), 
now 60 years ago.  They
are cousins of the more luminous quasars, having the advantage of
being much more numerous, so that they include many nearby objects
that can be studied at high linear resolution.  Seyfert galaxies exhibit
forbidden optical emission lines of moderate width (a few hundred km~s$^{-1}$),
while their permitted emission lines may be very broad (Seyfert~1
galaxies, with line widths of a few thousand km~s$^{-1}$) or of
widths similar to the forbidden lines (Seyfert~2 galaxies).
This has been explained generally by a unified scheme in which
Seyfert galaxies contain an obscuring nuclear torus or disk that
is seen pole-on in Seyfert 1 galaxies or edge-on in Seyfert 2
galaxies (e.g., Antonucci 1993).  Most of the observational
manifestations of Seyfert galaxies, certainly in the inner few
milliarcseconds, are thought to be powered
by central supermassive black holes that typically have
masses in the range of $\sim 10^7 M_{\sun}$ to $\sim 10^8 M_{\sun}$
(e.g., Wandel, Peterson, \& Malkan 1999).

\section{Scientific Goals and Context}

There are a number of scientific problems that can be addressed
by imaging Seyfert galaxies on milliarcsecond scales, a realm that
is just beginning to be explored in these objects.  Among the key 
questions are the following:

\begin{enumerate}

\item Why is there such a wide range of spectral properties in active galactic
nuclei (AGNs), with radio/optical luminosity ratios spanning many orders
of magnitude?

\item What do Seyfert galaxies reveal about the formation and fueling
of supermassive black holes and radio jets?

\item What are the observational manifestations of supermassive black
holes in nearby galaxies?

\end{enumerate}

The above questions can be asked in the context of several currently
unsolved problems in AGN physics.  Specifically, one current question
of interest is the assessment of the differences between radio-quiet
and radio-loud AGNs.  Are there two populations or a continuous 
distribution, and what are the parameters that control the
relative strengths in different spectral regimes?  Another question
is that of the general nature and physics of the low-luminosity AGNs
(LLAGNs).  Here, we would like to know how LLAGN accretion and 
jet-formation processes differ from radio galaxies and quasars.

The general areas of interest lead to a few specific observations
of Seyfert galaxies that may be made with VLBI arrays such as the
Very Long Baseline Array (VLBA):

\begin{enumerate}

\item {\it Jet properties and speeds:}  Are Seyfert and LLAGN jets 
intrinsically weak and slow, or quenched by gas in their nuclei?
Do VLA and VLBI-scale jets have the same symmetry axes?

\item {\it Compact radio-source spectra and luminosities:}  Are
these quantities consistent with various flavors of accretion models,
or jet models, or is a combination required?

\item {\it Radio core sizes:}  For nearby galaxies at distances of
tens of megaparsecs, classical models of advection-dominated accretion
flows, or ADAFs (e.g., Narayan, Mahadevan, \& Quataert 1998) predict sizes of 
tens of microarcseconds. Models
in which the radio emission is dominated by compact jets (e.g., Falcke
1999) predict sizes of hundreds to thousands of microarcseconds.
Can the two be distinguished by means of VLBI imaging?

\item {\it Location of gas relative to the AGNs:}  The location of 
various parcels of gas can be studied by imaging 
H$_2$O megamaser emission (Greenhill et al. 2003), 
free-free absorption (Walker et al. 2000), or HI
absorption (Mundell et al. 2003).  Discussion of this issue is 
beyond the scope of the current contribution, and the reader
is referred to the cited references instead.

\end{enumerate} 

\section{Jet Properties and Speeds}

A few Seyfert galaxy jets have been imaged at multiple epochs, so that
their jet speeds can be measured on parsec scales.  Examples are 
shown in Figure~1, taken from Ulvestad et al. (1999).  Here, the
cores of the Seyfert galaxies Mrk~231 and Mrk~348 are shown at
two epochs from early VLBA imaging; the respective relative
component speeds are $\beta_{\rm app} = 0.14\pm 0.052$ and
$0.074\pm 0.035$, where $\beta_{\rm app}\equiv v_{\rm app}/c$.  
These values appear fairly typical of most Seyfert
galaxies measured to date, although objects in outburst, such
as III~Zw~2 (Brunthaler et al. 2000) and Mrk~348 (Peck et al. 2003)
may display apparent speeds on sub-parsec scales that are
(at least) a substantial fraction of the speed of light.

\begin{figure}
\plottwo{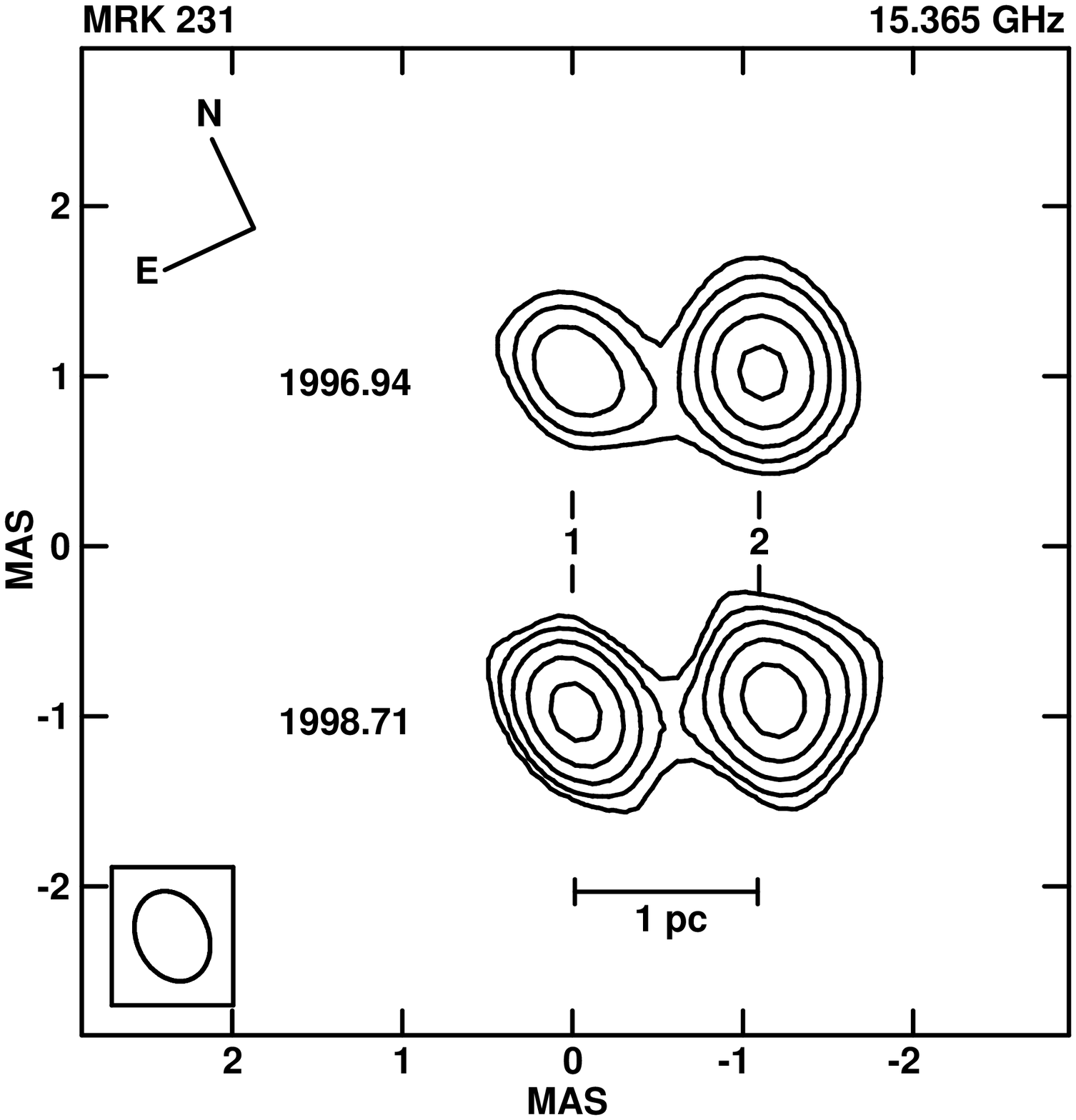}{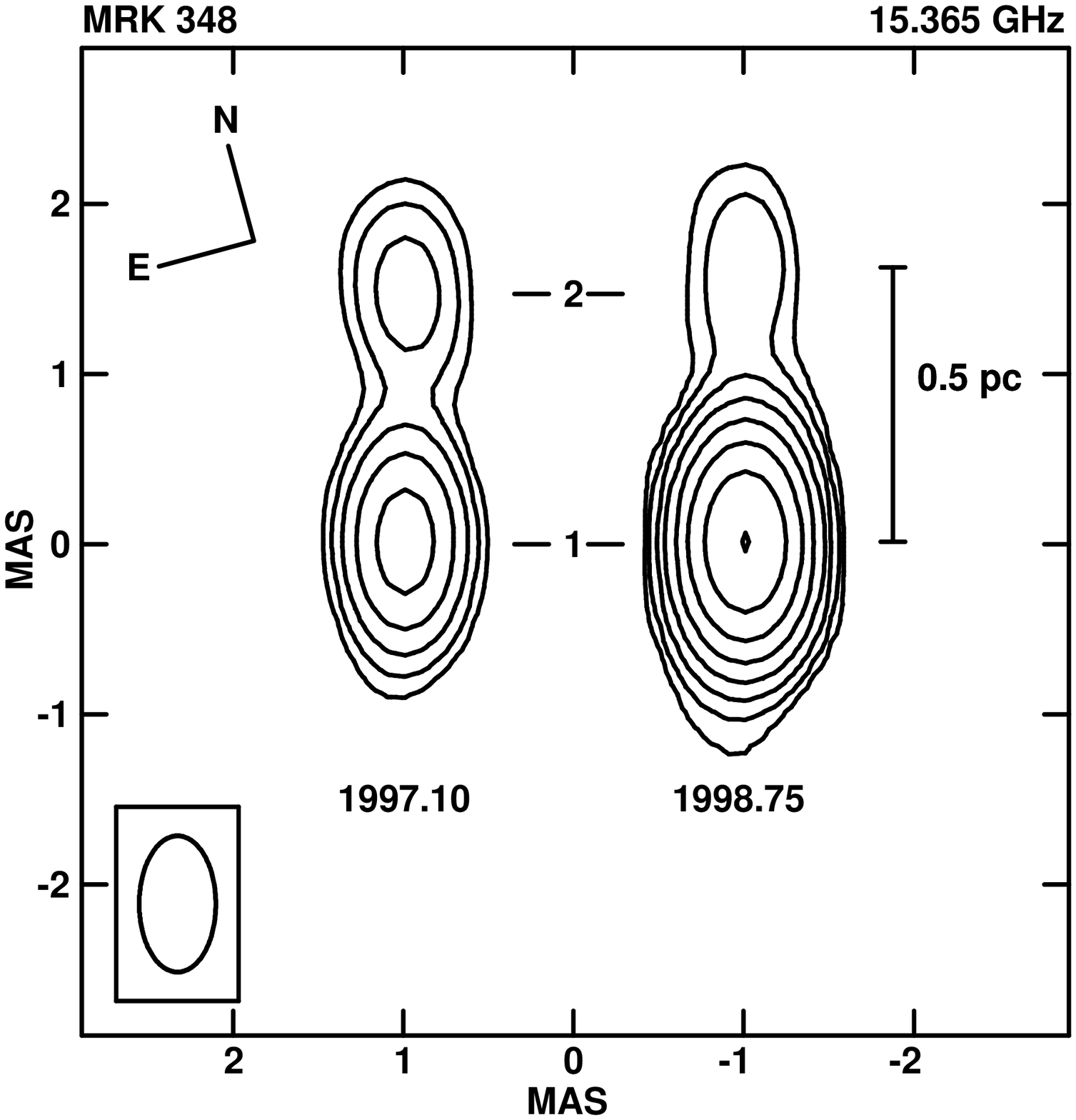}
\caption{Two-epoch VLBI images of the Seyfert galaxies
Mrk~231 ({\it left}) and Mrk~348 ({\it right}) are shown,
taken from Ulvestad et al. (1999).  The apparent speeds
corresponding to the small relative motions of the two
components are $\sim 0.1c$ within the central parsec of
each galaxy.}
\end{figure}

A summary of published measurements of Seyfert jet 
speeds is given in Table~1.  In contrast to the Seyfert 
galaxies, radio galaxies and radio-loud quasars generally have
highly relativistic jets on scales of a few parsecs to hundreds of
parsecs.  Since the properties of nuclear gas in Seyfert galaxies and
the more powerful objects are fairly similar (as indicated by
emission-line measurements), it seems likely that Seyfert galaxy jets 
initially are much less powerful relative to the nuclear
environment, and may even have different acceleration mechanisms.

\begin{table}
\caption{Published jet speeds in Seyfert galaxies}
\begin{tabular}{lcccl}
\tableline
Galaxy&$\beta_{\rm app}$&Scale&Comment&Reference \\
\tableline
III Zw 2&$<0.04$&0.2 pc&Quiescent&Brunthaler et al. (2000) \\
&$>1.25$&0.4 pc&Outburst&Brunthaler et al. (2000) \\
Mrk 348&0.07&0.5 pc&Start of outburst&Ulvestad et al. (1999) \\
& $\sim 0.4$&0.5 pc&Outburst&Peck et al. (2003) \\
NGC 1068&$<0.08$&21 pc&Quiescent&Roy et al. (2000) \\
NGC 4151&$<0.14$&7 pc&Quiescent&Ulvestad et al. (1998) \\
&$<0.05$&7 pc&Quiescent&Ulvestad et al., in prep. \\
Mrk 231&0.14&1 pc&Moderate outburst&Ulvestad et al. (1999) \\
NGC 5506&$<0.25$&3.4 pc&Quiescent?&Roy et al. (2001) \\
\tableline
\tableline
\end{tabular}
\end{table}

At present, there is no obvious systematic difference
between objects of different Seyfert classes in the few objects
studied. In a large sample, we might expect that Seyfert 2 galaxies 
would have larger apparent jet speeds as a class; their more edge-on
tori should translate into jets that have a larger projection
in the plane of the sky than for Seyfert 1 galaxies.  However,
this would be true only if the small-scale jet axes
truly indicate the orientation of the nuclear torus.

Recently, it has been shown conclusively that VLA-scale jets in
Seyfert galaxies essentially are randomly oriented with respect
to their host galaxy major axes (Kinney et al. 2000). This generally
is thought to indicate that the central tori are misaligned with
respect to the galaxy disks, and therefore had a different origin.  
However, it is not necessarily the
case that the VLA-scale jet must indicate the direction of the 
torus axis, since there often is significant curvature between
milliarcsecond and arcsecond scales.  For example, Mrk~231
shows a position angle difference of 65\deg\ between scales of
1~pc and tens of parsecs (Ulvestad, Wrobel, \& Carilli 1999) and
NGC~5506 has a jet bend of 90\deg\ at a few parsecs from the nucleus
(Roy et al. 2001).  The beautiful 1.4-GHz continuum image of 
NGC~4151 by Mundell et al. (2003) demonstrates that
a jet which looks straight in VLA and MERLIN images (e.g., 
Mundell et al. 1994) actually
contains a number of remarkable changes in jet direction
when viewed with milliarcsecond resolution, possibly due to 
interactions with narrow-line gas clouds.  Given that 
nuclear tori can be strongly warped by radiation
on sub-parsec scales (Pringle 1997), we must be wary of
inferring the directions of the central tori even from 
milliarcsecond-scale jet directions.

\section{Compact Radio-source Spectra and Luminosities}  

Significant samples of LLAGNs, including both Seyfert and LINER
galaxies, have been imaged with the VLBA in order to determine
whether they contain milliarcsecond-scale cores that must be
powered by supermassive black holes (Falcke et al. 2000;
Nagar et al. 2002).  In virtually all objects that have unresolved
cores on arcsecond scales, a compact radio source with a
brightness temperature $T_{\rm B}\geq 10^8$~K has been detected,
a sure indicator of a nucleus powered by a massive black hole.

Classical Seyfert galaxies typically have steep radio spectra,
indicative of optically thin synchrotron radiation; no more than
$\sim 10$\% of such objects appear to have flat or inverted spectra
(Ulvestad \& Wilson 1989).  In the classical Seyfert galaxy NGC~1068,
a weak flat-spectrum radio core has been resolved by the VLBA. 
This core has multiple components with brightness temperatures of
$\sim 10^6$~K, interpreted as direct or reflected emission from
the nuclear torus itself (Gallimore, Baum, \& O'Dea 1997).  
However, efforts to find similar objects by VLBA imaging of 
flat-spectrum cores in other classical Seyfert galaxies have failed 
thus far (Wilson et al. 1998; Mundell et al. 2000), indicating that 
most of these objects are likely to have flat spectra due to 
synchrotron self-absorption.  In the classical Seyfert NGC~4151,
a high-sensitivity VLBI observation in 2002, using the VLBA together with
several 100-m class telescopes, has revealed a flat-spectrum
2-mJy radio core at 15~GHz. That core is shown here in Figure~2; 
it appears to be the AGN and the origin of the two-sided 
jet seen on larger scales, but shows no evidence of emission from 
a nuclear torus (Ulvestad et al. 2003, in preparation).  Comparison of 
this image with an 8.4~GHz image made from data taken in 1998 is
under way, in order to measure the apparent speed on a 0.1-pc scale.

\begin{figure}
\plotone{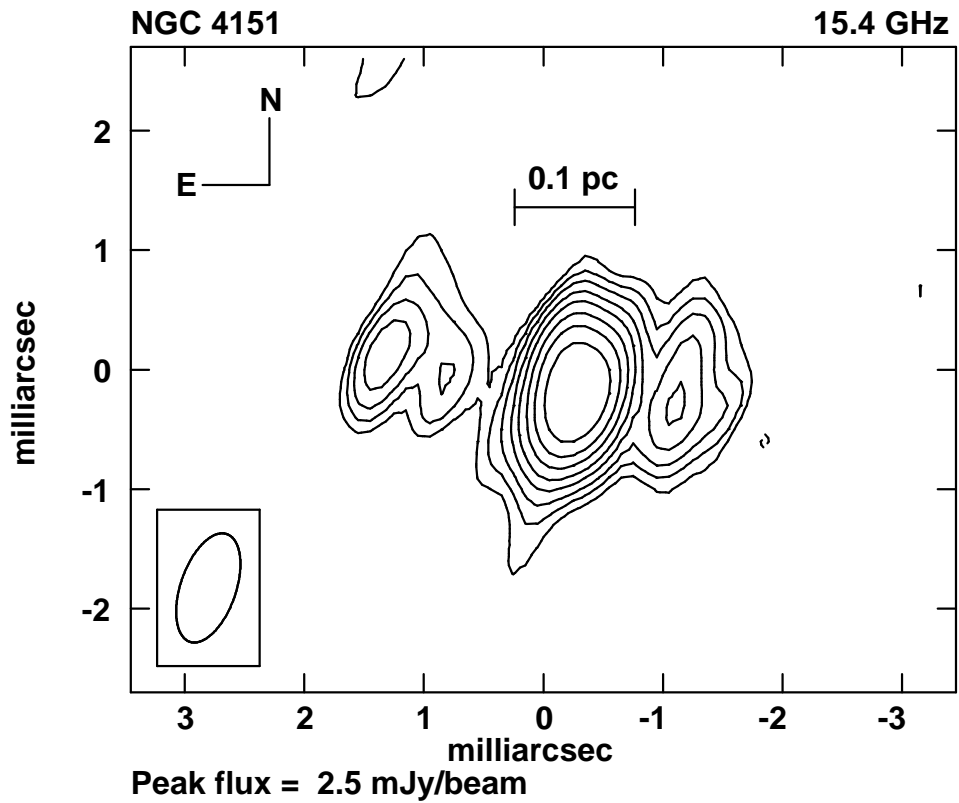}
\caption{VLBI image of the AGN in NGC~4151, made using the
VLBA, phased VLA, Green Bank Telescope, and Effelsberg at
15 GHz (Ulvestad et al. 2003, in prep.).  Contour levels 
are incremented by factors of $2^{1/2}$, beginning at
3 times the rms noise of 45~$\mu$Jy~beam$^{-1}$.}
\end{figure}

Surprisingly, within the lower-luminosity Palomar 
Seyfert sample (Ho, Filippenko, \& Sargent 1997), 
it appears that roughly half of the objects in an
optically selected sample have flat or inverted radio spectra
(Ulvestad \& Ho 2001a).  In these objects, which appear to
be radiating far below their Eddington luminosities, the
accretion flows almost certainly have low enough densities
so that they have low radiative efficiencies,  Then,
synchrotron radiation from hot electrons in the accreting
gas can be self-absorbed, yielding a slightly inverted spectrum 
at radio frequencies in the gigahertz range (Narayan et al. 1998).  
Ulvestad \& Ho (2001b) used the VLBA to image three of the stronger 
LLAGNs with inverted spectra at radio frequencies ranging from 
1.6~GHz to 8.4~GHz.  In all three cases, they found unresolved
sources with radio spectra that increase gradually with frequency, 
similar to the spectrum of the radio source Sgr~A* at the center 
of our own galaxy.

One of the well-known problems of classical ADAF models is that
the radio emission in LLAGNs and Sgr~A* is observed to be 
far stronger relative to the X-ray emission than is predicted
by the models.  This problem can be ``fixed'' if a large fraction
of the observed radio emission comes from a compact radio jet rather than
from the accretion flow itself (Falcke \& Markoff 2000; Falcke 2001).
The jet models seem to predict that the radio emission
should become optically thin and turn over above a few gigahertz.
However, Anderson \& Ulvestad (2003, in preparation) have used the VLBA
to image the LLAGNs from Ulvestad \& Ho (2001b) at frequencies up
to 43~GHz, and found that the radio spectra continue to be flat or
rising with frequency, in contradiction to the current jet models.

Ho \& Peng (2001) and Terashima \& Wilson (2003) have used high-resolution
optical and X-ray imaging to attempt to isolate the active nuclei of
Seyfert galaxies and LLAGNs, and to divide them into radio-quiet and radio-loud
objects.  Terashima \& Wilson studied the radio/X-ray ratio using the 
quantity $R_X\equiv \nu L_\nu({\rm 5\ GHz})/L_X(2-10\ {\rm keV})$,
useful even for heavily obscured nuclei.  They found that
$\log R_X=-4.5$ is the rough dividing line between radio-quiet
and radio-loud objects. When compact radio flux is used, most of 
their Seyfert galaxies have $\log R_X\sim -5$ and are in the
radio-quiet category, whereas using the total radio flux would make
more of these objects radio loud.  The general trend found by
Terashima \& Wilson (2003), using the VLBI core powers from some of the
references cited above, is that most LLAGNs are radio
loud.  These LLAGNs almost certainly have accretion flows with
low radiative efficiency. Therefore, the implication is that the central 
power sources in such objects put a significantly larger fraction of their 
total energy output into radio emission than in galaxies with 
classical accretion flows.

\section{Radio Core Sizes}

Radio emission from low-efficiency accretion flows is thought to
originate very close to the central black holes, perhaps within a 
few to a few tens of Schwarzschild radii (Mahadevan 1997).  In
contrast, emission from compact jets connected to these flows
is thought to come from scales a factor of $\sim$30--40 times
larger, up to as much as 1000 Schwarzschild radii (Falcke \&
Biermann 1999).  Ulvestad \& Ho (2001b)
found upper limits of $10^3$--$10^4$ Schwarzschild radii from
their VLBA imaging of three LLAGNs, and these limits have been
extended downward by factors of $\sim 5$ by 43-GHz VLBA imaging
of the same objects (Anderson \& Ulvestad 2003, in preparation).
The upper limits now are getting uncomfortably small for the
compact jet models.  Putting this together with the problem mentioned 
above, that the radio spectra do not turn over above $\sim 10$~GHz,
it is becoming clear that the accretion/jet models used to date
may need more work in order to account for the compact radio 
emission from LLAGNs.

Since the Earth is not going to get any bigger, what can we do
in order to attempt to measure true sizes of the radio sources in
low-luminosity Seyferts and AGNs?  Space VLBI is one option, but
the relative weakness of the radio cores in most Seyfert galaxies
makes this only a fond hope for the relatively distant future.
But a more promising technique may be to use the interstellar
medium in our Galaxy to ``image'' the radio cores.  It recently has
become clear that intraday variability in strong extragalactic
radio sources is caused by scintillation in the interstellar medium
(Rickett et al. 2001; Jauncey \& Macquart 2001; Dennett-Thorpe \& de 
Bruyn 2002), providing a means of estimating radio core sizes in the 
vicinity of a few microarcseconds.  Although measuring intraday 
variability in millijansky-strength sources will be a real challenge to 
modern instrumentation, it may be possible with the current VLA,
and certainly is well within the reach of the more sensitive
Expanded VLA.

\medskip

\noindent{\bf Acknowledgments.} The National Radio Astronomy Observatory
is a facility of the National Science Foundation operated under
cooperative agreement by Associated Universities, Inc.  I thank
the many collaborators whose work has been summarized in this
contribution.

\end{document}